\documentclass[12pt]{article}

\usepackage{cancel}
\usepackage{amsfonts,amsmath}
\usepackage{epstopdf,amsfonts, amsmath,amssymb}
\usepackage{stmaryrd}
\usepackage{tikz}
\usetikzlibrary{decorations.markings}
\numberwithin{equation}{section}

\numberwithin{equation}{section}

\newcommand\encadremath[1]{\vbox{\hrule\hbox{\vrule\kern8pt
\vbox{\kern8pt \hbox{$\displaystyle #1$}\kern8pt}
\kern8pt\vrule}\hrule}}
\def\enca#1{\vbox{\hrule\hbox{
\vrule\kern8pt\vbox{\kern8pt \hbox{$\displaystyle #1$}
\kern8pt} \kern8pt\vrule}\hrule}}

\newcommand\figureframex[3]{
\begin{figure}[bth]
\hrule\hbox{\vrule\kern8pt
\vbox{\kern8pt \vbox{
\begin{center}
{\mbox{\epsfxsize =#1.truecm\epsfbox{#2}}}
\end{center}
\caption{#3}
}\kern8pt}
\kern8pt\vrule}\hrule
\end{figure}
}
\newcommand\figureframey[3]{
\begin{figure}[bth]
\hrule\hbox{\vrule\kern8pt
\vbox{\kern8pt \vbox{
\begin{center}
{\mbox{\epsfysize =#1.truecm\epsfbox{#2}}}
\end{center}
\caption{#3}
}\kern8pt}
\kern8pt\vrule}\hrule
\end{figure}
}

\newtheorem{theorem}{Theorem}[section]

\newtheorem{remark}[theorem]{Remark}
\newtheorem{proposition}[theorem]{Proposition}
\newtheorem{definition}[theorem]{Definition}
\newtheorem{lemma}[theorem]{Lemma}
\newtheorem{corollary}[theorem]{Corollary}
\def\br{\begin{remark}\rm\small}
\def\er{\end{remark}}
\def\bt{\begin{theorem}}
\def\et{\end{theorem}}
\def\bd{\begin{definition}}
\def\ed{\end{definition}}
\def\bp{\begin{proposition}}
\def\ep{\end{proposition}}
\def\bl{\begin{lemma}}
\def\el{\end{lemma}}
\def\bc{\begin{corollary}}
\def\ec{\end{corollary}}
\def\beaq{\begin{eqnarray}}
\def\eeaq{\end{eqnarray}}

\newcommand{\beq}{\begin{equation}}
\newcommand{\eeq}{\end{equation}}
\newcommand{\bea}{\begin{eqnarray}}
\newcommand{\eea}{\end{eqnarray}}

\newcommand{\ad}{\mathrm{ad}\,}

\newcommand{\diff}{\text{d}}
\newcommand{\Ho}{\operatorname{H}}

\newcommand{\Lieg}{{\mathfrak g}}

\newcommand{\Ad}{\operatorname{Ad}}
\newcommand{\Ker}{\operatorname{Ker}}
\newcommand{\Img}{\operatorname{Im}}

\DeclareUnicodeCharacter{202F}{\,}

\usepackage[pdftex]{hyperref}
\hypersetup{colorlinks, urlcolor=magenta,citecolor =blue, linkcolor=blue, filecolor=black}

\begin{document}

\sloppy

\addtolength{\baselineskip}{0.20\baselineskip}
\begin{center}
\vspace{2cm}

{\Large \bf {A renormalised description of meromorphic connections over curves}% \\ \large \bf{and its semi-classical counter-part}
}

\vspace{1cm}

\begin{large}{Rapha\"el Belliard}\end{large}

%\vspace{5mm}
\begin{small}
Department of Mathematical and Statistical Sciences, \\ University of Alberta, Edmonton, AB T6G 2G1 Canada \\ \texttt{belliard@ualberta.ca}
\end{small}

\end{center}

\vspace{0.01cm}
\begin{center}
{\bf Abstract}
\end{center}

\begin{small}
The Renormalisation Group (RG) is a systematic procedure used to regularise divergences appearing as artefacts when constructing solutions to a large class of differential problems, whether perturbatively or not. This paper is devoted to performing it in the non-perturbative context of flat sections of meromorphic connections in holomorphic principal bundles over a base complex curve, with reductive and complex structure Lie groups. Using the interpretation of Kunuhiro of RG in terms of envelopes of families of solutions, we identify these envelopes as particular flat sections associated to points on the corresponding quantum spectral curve, namely certain divisors on the base curve with local coefficients in the adjoint bundle, yielding integral representations. This shows that the non-perturbative renormalisation of flat sections of principal bundles over curves fits naturally within the description of meromorphic connections via their quantum geometry, or homology \textit{\`a la} Goldman, where the flat sections and their deformations are parameterized by degree zero and one chains respectively. We conclude the paper by mentioning upcoming applications.

%\pagebreak

%\vspace{5mm}
\tableofcontents 
\end{small}

\pagebreak

\section{Introduction}
\label{Intro}

Families of Lie-equivariant meromorphic connections in principal bundles over curves have become ubiquitous in mathematical physics. They appear for instance as constitutive geometry of certain integrable systems \cite{Boa2002, ArRe2021}, and as quantum curves in the perturbative contexts of the Topological Recursion \cite{EyOr2007, KoSo2017} and Wentzel-Brillouin-Kramers (WKB) analysis \cite{Dun1932, BoEy2017, EGFMO2021}.

The present work describes certain structural results eventually aimed at bridging between perturbative and non-perturbative uses of meromorphic connections. In order to do so, we work out the interplay between a contemporary twist of Goldman's homology with local coefficients \cite{Gol1984, BER2018, BeEy2019} and the modern description of the RG procedure through the classical theory of envelopes \cite{CGO199X, Kun1995}.

\subsection{Setup}

In this paper, we will consider geometric structures over a fixed smooth and compact complex curve, or \textit{Riemann surface}, denoted $\mathcal C$ \cite{FaKr1980}. These geometric structures will additionally depend on a set of distinct points on the curve, $\bold p=\{p_1,\dots,p_M\}\subset\mathcal C$, for some integer $M\in\mathbb Z_{\geq0}$, as well as on a reductive complex Lie group $G$ \cite{Hum1980}. Let us denote by $\mathcal C_{\bold p}=\mathcal C-\bold p$ the corresponding punctured curve.

\subsection{Organisation of the paper}

The paper is organised as follows:
\begin{itemize}

\item In Section \ref{geometry}, we recall the definitions of holomorphic principal $G$-bundles over $\mathcal C$, that of meromorphic connections thereof, that of the corresponding quantum geometry, namely the extension of Goldman's homology with local coefficients \cite{Gol1984} that was introduced in \cite{BER2018, BeEy2019}, concluding with how they relate to describe deformations of the setup at fixed complex structures.

\item In Section \ref{renormalisation}, after introducing Kunuhiro's formulation of RG in terms of the theory of envelopes \cite{Kun1995}, we show that its application to flat section equations in principal bundles is equivalent to the definition of solutions by using regularised pairings with generalised cycles.

\item In Section \ref{conclusions}, we gather the content of the paper in the form of the homological parameterization of meromorphic connections and their deformations, dual to the natural one in terms of \v{C}ech cohomology. We then conclude the paper by mentioning some upcoming work under development.

\end{itemize}

\section{Quantum geometry}
\label{geometry}

\subsection{$G$-bundles with meromorphic connections}
\label{principal}

For the present purposes, we consider a fixed holomorphic principal $G$-bundle $\mathcal P\longrightarrow\mathcal C$ of degree zero over the base curve $\mathcal C$. It is a fibre-bundle with generic fibre isomorphic to the structure group $G$, determined by the corresponding collection of \textit{holomorphic transition functions}, whose orbit under bundle-isomorphisms defines a class in the first \v{C}ech cohomology space $\Ho^1(\mathcal C,\mathcal O_{\mathcal C}\otimes G)$. We accordingly denote the corresponding adjoint bundle by $\ad\mathcal P\longrightarrow\mathcal C$, whose generic fibre is isomorphic to the Lie algebra $\Lieg:=\text{Lie}(G)$ of $G$, and whose transition functions are identical to that of $\mathcal P$, albeit acting in the adjoint representation.

The degree zero assumption implies that the bundles under consideration is topologically trivial but not holomorphically so. Nevertheless, although we will be interested in deformations of certain geometric structures related to those bundles, we will solely consider those leaving the underlying complex structure invariant. 

In particular, we consider families of meromorphic $G$-connections in $\mathcal P$, generically denoted $\nabla$, defined as bundle maps $\nabla:\mathcal P\longrightarrow\mathcal P\otimes  K(D)$, $ K\longrightarrow \mathcal C$ denoting the \textit{canonical bundle} of holomorphic $(1,0)$-forms on $\mathcal C$, $D$ the divisor of poles of $\nabla$, such that it satisfies
\begin{eqnarray}
\label{Leibniz}
\nabla(\lambda \Psi)\, =\, (\diff\lambda)\Psi + \lambda\nabla\Psi
\end{eqnarray}
for all choices of local holomorphic function $\lambda$ and section $\Psi$, on the interior of the intersection of their definition sets. \textit{Any} reference connection allows to trivialise the bundle in which it is defined, and can be used as a reference point to parameterize meromorphic connections thereof. Canonical \textit{holomorphic} candidates are for instance provided in \cite{BiHu2021}, and we will consider the bundle trivialised, from now on writing the connection under consideration as $\nabla=\diff-\Phi$, with adjoint-valued gauge potential $\Phi$. We refer the reader to \cite{Boa2002, KaWe2017, Boa2020} for global and local descriptions of moduli of meromorphic connections in $G$-bundles over curves.

\subsection{Homology with local coefficients}
\label{homology}

The homology with local coefficients considered in \cite{BER2018, BeEy2019} is a direct extension of that of \cite{Gol1984} to the case of meromorphic connections. It consists in the sequence of homology groups associated to the complex $\big(\mathfrak C_\bullet, \widehat\partial_\bullet\big)$ of chains $\gamma^{(k)}\otimes\sigma\in\mathfrak C_k$, $k\geq0$, $\gamma^{(k)}:[0,1]^k\longrightarrow\mathcal C$ being a continuous map and $\sigma$ denoting an $\nabla$-flat section of $\ad\mathcal P$ defined in a neighbourhood of $\gamma^{(k)}([0,1]^k)$ from which the singularities of $\nabla$ have been excluded, and together with the graded boundary operators defined by
\begin{eqnarray}
\label{boundary}
\widehat\partial_k\big( \gamma^{(k)}\otimes\sigma\big)\, := \, \sum_i \pm(\partial\gamma^{(k)})_i\otimes \sigma\big|_{(\partial\gamma^{(k)})_i}.
\end{eqnarray}
The sum runs over the boundary components $(\partial\gamma^{(k)})_i$ of the image of $\gamma^{(k)}$ that do not intersect the divisor of singularities of $\nabla$, the $\pm$ sign accounting for relative orientation. We will hereafter drop the tensor product symbol when no confusion is possible. The corresponding homology groups,  whose elements are called \textit{generalised cycles}, are then defined by
\begin{eqnarray}
\widehat \Ho_k \, := \, \Ker \widehat \partial_k \big/ \Img \widehat \partial_{k+1}.
\end{eqnarray}
This definition is equivalent to the requirement that pairings with Lie-algebra valued holomorphic differentials are well-defined, in the sense that they depend only on the homology classes of chains representing such generalised cycles.

We refer to \cite{Gol1984, BER2018, BeEy2019} for corresponding in-depth descriptions but let us make a couple of remarks before carrying on. Firstly, the higher homology groups $\widehat \Ho_{k\geq2}=\{0\}$ are all trivial.

Secondly, the vanishing boundary conditions defining the elements of $\widehat \Ho_1$ become collections of Lie-algebra valued relations between the involved sections, when evaluated at the underlying boundary divisor defined by (\ref{boundary}). The simplest of these elements, called \textit{trivial cycles} in a sense we precise in the coming subsection, are obtained from chains of the form $\gamma^{(1)}\otimes \sigma$, where $\gamma^{(1)}$ draws a \textit{small} simple circle around a pole of $\nabla$. Its boundary then takes the form
\begin{eqnarray}
\label{trivial}
\widehat\partial_1\big(\gamma^{(1)}\otimes\sigma\big)\, = \, \gamma^{(1)}(0)\, \big\{ \sigma\big(\gamma^{(1)}(1)\big) -\sigma\big(\gamma^{(1)}(0)\big)\big\},
\end{eqnarray}
and its vanishing simply becomes the requirement that its value at $\gamma^{(1)}(0)$ is parallel transported to itself along $\gamma^{(1)}$. In other words, that value has to commute with the corresponding monodromy matrix of the connection. Finally, $\widehat\Ho_0$ is trivial.

In \cite{BER2018, BeEy2019}, the total space of the adjoint bundle was coined \textit{quantum spectral curve}, justified by the vanishing of all higher homology groups. For the rest of this paper, we will call the elements of $\mathfrak C_1$ \textit{quantum trajectories}, moreover \textit{closed} for underlying classes belonging to $\widehat\Ho_1$, and \textit{open} otherwise. Furthermore, we will call \textit{quantum divisors} the elements of $\mathfrak C_0$, whose boundaries necessarily vanish. The \textit{support} of a quantum divisor $X=\sum_{x\in\mathcal C}x E_x\in\mathfrak C_0$ is defined as the \textit{finite} collection of base-points $x\in\mathcal C$ with non-trivial corresponding adjoint elements $E_x\neq0$. Correspondingly, \textit{localised} quantum divisors are defined to be those whose support consists of a unique point, the collection of which we denote $\mathfrak C_0'$.

\pagebreak

\subsection{Deformation families}
\label{deformations}

In \cite{BER2018, BeEy2019}, it was shown that the elements of the first homology space $\widehat\Ho_1$ that are \textit{not} trivial in the sense of the previous subsection are in one-to-one correspondence with deformations of the meromorphic connection under consideration (keeping the underlying complex structure fixed). More precisely, for every cycle $\Gamma\in\widehat\Ho_1$, one can consider the adjoint-valued function
\begin{eqnarray}
\label{DeformationPotential}
F_\Gamma(x) & := & \oint_\Gamma \omega_{x,o}\\
&=& \sum_i \int_{y\in\gamma_i} \omega_{x,o}(y)\sigma_i(y),
\end{eqnarray}
where the cycle was represented in the form $\Gamma = \sum_i \gamma_i \otimes \sigma_i$, and $\omega_{x,o}$ is \textit{any} fixed meromorphic one-form on $\mathcal C$ with a simple pole at $x\in\mathcal C_{\bold p}$ with residue $+1$, a simple pole at a reference point $o\in\mathcal C_{\bold p}$ with residue $-1$, and no other singularities. It follows that we can define a deformation $\partial_\Gamma\Phi$ of the gauge potential as
\begin{eqnarray}
\partial_\Gamma\Phi\, := \, \diff F_\Gamma + [F_\Gamma,\Phi].
\end{eqnarray}
The trivial cycles of the previous subsection are then precisely those corresponding to trivial deformations through that construction. 

The rest of this paper consists in completing the picture arising through this duality between closed quantum trajectories and deformations of meromorphic connections, by identifying their flat adjoint sections with localised quantum divisors. 

The reader unfamiliar with the present notion of quantum geometry might wonder at this point how the Poincar\'e pairing featured in (\ref{DeformationPotential}) is supposed to be performed when the projection of the underlying closed quantum trajectory to the base curve intersects the singular locus of the connection. In \cite{BER2018, BeEy2019}, we introduced \textit{ad hoc} regularisations consisting in the subtraction of the singular parts appearing in the definitions of the underlying improper integrals. 

In the coming section, we start by recalling the notion of envelopes of families of curves, underlying the process from which our simple identification will emerge.  However before applying it to solutions of the flat section equation, we will illustrate it by showing how it gives a natural geometric interpretation to the aforementioned regularisation of divergent integrals.

\section{Renormalisation}
\label{renormalisation}

\subsection{Envelopes of families of curves}
\label{envelopes}

In \cite{Kun1995}, the Renormalisation Group (RG) procedure from theoretical physics \cite{GeLo1954, Cal1970, Sym1970, WiFi1972}, in its modern presentation \cite{CGO199X}, was formulated as the process of taking envelopes of one-parameter families of solutions of the underlying (differential) problems, or families of curves in the relevant solution spaces. We now briefly recall how such envelopes are determined, as can for instance be found in the classical text \cite{CoHi1962}.

Let us therefore consider an analytic family $\text m_\bullet=\{\text m_\tau\}_{\tau\in\mathcal T}$ labelled by points in a fixed parameter space $\mathcal T$ of complex dimension $1$, where for each $\tau\in \mathcal T$, $\text m_\tau$ is a map from $\mathcal T$ to a fixed vector space $\mathbb V$, therefore drawing a curve $\{\text m_\tau(t)\}_{t\in\mathcal T}$ in $\mathbb V$. The envelope $\overline{\text m}=\{\overline{\text m}(t)\}_{t\in\mathcal T}$ of the family $\text m_\bullet$ is then defined as the curve in $\mathbb V$ given by
\begin{eqnarray}
\label{Envelope}
\overline{\text m}(t)\, := \, \text m_{\tau_*(t)}(t),
\end{eqnarray}
where for all $t\in\mathcal T$, $\tau_*(t)$ is defined by the \textit{envelope condition} 
\begin{eqnarray}
\label{EnvelopeCondition}
\frac\partial{\partial \tau}\text m_\tau(t)\Big|_{\tau_*(t)}\, = \, 0.
\end{eqnarray}

We will apply these ideas to flat sections of the trivialised adjoint bundle, namely $\ad\mathcal P\simeq \mathcal C_{\bold p}\times \Lieg$, labelled by locations of initial conditions in the form of integration constants, and where we identify the underlying vector space to be $\mathbb V=\Lieg$. We will interpret them in terms of boundaries of open quantum trajectories but before getting to that, let us illustrate how the envelope point of view on RG allows us to give a geometric description of the divergent integral regularisation mentioned in last section.

\subsection{Renormalised divergent integrals}
\label{DivergentIntegrals}

The purpose of this paragraph is two-fold. Firstly, recall how divergent complex integrals of the form \begin{eqnarray}
\int_p^z f(\zeta) \frac{\diff\zeta}{(\zeta-p)^{d+1}}\, = \, \infty,
\end{eqnarray}
for some distinct complex numbers $p\neq z \in\mathbb C$, some Lie-algebra valued function $f$, analytic on a domain containing both $z$ and $p$, and some non-negative integer $d\geq0$, were regularised in \cite{BER2018, BeEy2019}. Secondly, see how these regularisations in fact consist in performing RG in the sense that their output satisfy envelope conditions in the sense of (\ref{EnvelopeCondition}). Indeed, introducing the order $d$ Taylor asymptotics 
\begin{eqnarray}
f(\zeta) & \underset{\zeta\sim p} \sim & \sum_{k=0}^d \frac{f^{(k)}(p)}{k!} (\zeta-p)^k
\end{eqnarray}
of the function $f$ around $p$, the corresponding regularised integral is defined as
\begin{eqnarray}
\int_{[p,z]} f(\zeta) \frac{\diff\zeta}{(\zeta-p)^{d+1}} & := & \int_p^z \Big(f(\zeta)-\sum_{k=0}^d \frac{f^{(k)}(p)}{k!} (\zeta-p)^k \Big)\frac{\diff\zeta}{(\zeta-p)^{d+1}} \nonumber \\
&&\  -\ \ \sum_{k=0}^{d-1} \frac{f^{(k)}(p)}{k!(d-k)} \frac1{(z-p)^{d-k}} \, + \, \frac{f^{(d)}(p)}{d!} \ln (z-p).
\end{eqnarray}
This is now easily interpreted as the envelope of a family of curves by making the observation that
\begin{eqnarray}
\int_{[p,z]} f(\zeta) \frac{\diff\zeta}{(\zeta-p)^{d+1}} & = & \underset{q=p}\lim \  \Big\{ \int_q^z f(\zeta) \frac{\diff\zeta}{(\zeta-p)^{d+1}}\, -\, \sum_{k=0}^{d-1} \frac{f^{(k)}(p)}{k!(d-k)} \frac1{(q-p)^{d-k}} \nonumber \\
&&\qquad \qquad  +\ \ \frac{f^{(d)}(p)}{d!} \ln (q-p)\, + \, \frac{f^{(d+1)}(p)}{(d+1)!}(q-p)\Big\}.
\end{eqnarray}
The expression between curly brackets defines a family of primitives of the original integrand labelled by complex numbers $q$, that furthermore satisfies the envelope condition (\ref{EnvelopeCondition}) with respect to its label precisely when it coincides with the singular point. Indeed, taking the derivative of this expression with respect to the family label $q$, we find that it has the asymptotics 
\begin{eqnarray}
\Big(-f(q)+\sum_{k=0}^{d+1} \frac{f^{(k)}(p)}{k!} (q-p)^k \Big)\Big/(q-p)^{d+1} \ \underset{q\sim p}\sim \ - \sum_{k=d+2}^\infty \frac{f^{(k)}(p)}{k!} (q-p)^{k-d-1},
\end{eqnarray}
going to zero in the limit $q\rightarrow p$, and therefore realising the regularised integral as a renormalised envelope.

\subsection{Integral representations of flat sections}
\label{solutions}

In order to obtain integral representations of the flat sections through RG, we now make the working hypothesis that we have a generic solution $\Psi$ of the differential system, namely a \textit{multi-valued} section $\Psi:\mathcal C\longrightarrow\mathcal P$, see for instance \cite{Mag1954, IsNo1999}. This is not a light hypothesis and we will mention how it fits in our broader program in Section \ref{outlook}. It satisfies $\nabla\Psi=0$ by definition such that for every $E\in\Lieg$, the multi-valued adjoint section defined by $M_E:=\Ad_\Psi E$ satisfies formally the same equation $\nabla M_E=0$, albeit as a section of the adjoint bundle, namely
\begin{eqnarray}
\label{AdjointEquation}
\diff M_E \, = \, [\Phi, M_E]
\end{eqnarray}
in the trivialisation under consideration. We call $M_\bullet$ a \textit{fundamental adjoint solution}, and it allows to rewrite the solutions of the adjoint flat-section equation in the general parametric form
\begin{eqnarray}
\text m(zE;\zeta)\, := \, \int_{\zeta}^z \diff M_E + F(\zeta), 
\end{eqnarray}
satisfying the initial-value condition $\text m(\zeta E,\zeta)=F(\zeta)$, for any $E\in\Lieg$ and $\Lieg$-valued function $F$ encoding possible choices of initial value.

Following \cite{CoHi1962, Kun1995}, we get another solution of the differential system (\ref{AdjointEquation}) by considering the envelope $\overline{M_E}$ defined by
\begin{eqnarray}
\overline{M_E}(z)\, := \, \text m (z E; \zeta_*(z)),
\end{eqnarray}
where $\zeta_*(z)$ is required to satisfy the envelope condition modelled on (\ref{EnvelopeCondition}). Namely,
\begin{eqnarray}
\diff_\zeta \text m(zE;\zeta)\Big|_{\zeta_*(t)} = \, -\diff M_E(\zeta_*(z)) +\diff F(\zeta_*(z))\, = \, 0.
\end{eqnarray}
This constraint is solved in a straightforward way by taking
\begin{eqnarray}
\label{CounterTerm}
F(\zeta) \, := \, \int_{yH\in\Gamma(\zeta)} \diff M_H(y) \, = \, \int_{yH\in\Gamma(\zeta)} [\Phi(y), M_H(y)],
\end{eqnarray}
where $\Gamma(\zeta)\in\mathfrak C_1$ is any quantum trajectory with localised boundary
\begin{eqnarray}
\widehat\partial_1\Gamma(\zeta) \, =\, \zeta E \,\in\, \mathfrak C_0'
\end{eqnarray}
supported at $\zeta\in\mathcal C_{\bold p}$, by definition of the generalised boundary operator $\widehat\partial_1$. We also used the short-hand notation $yH\in\Gamma(z)$ to denote when $y\in\gamma$ with $\gamma\otimes M_H$ being a component of $\Gamma(z)$. 

The reader might again wonder about the convergence of such integrals over the singular locus of the connection, where the fundamental adjoint solution $M_\bullet$ is expected to exhibit essential singularities. However, and as was described in \cite{BeEy2019} by a process called \textit{squeezing}, we can assume without loss of generality that for any component $\gamma\otimes M_H$ of $\Gamma(z)$ such that $p\in\gamma([0,1])$ for some $p\in \bold p$, the Lie algebra element $H\in\Lieg$ is such that the singularity of $M_H(y)$ at $y=p$ is in fact a pole, and hence that the regularisation process of Section \ref{DivergentIntegrals} applies to (\ref{CounterTerm}).

Concatenating, we eventually obtain integral representations for the family of envelope solutions that are given by
\begin{eqnarray}
\label{IntegralForm}
\overline{M_E}(z) \, = \, \int_{yF\in\Gamma(z)} [\Phi(y),M_F(y)],
\end{eqnarray}
for \textit{any} $E\in\Lieg$ and quantum trajectory $\Gamma(z)\in\mathfrak C_1$ whose boundary is the localised quantum divisor given by $\widehat \partial_1 \Gamma (z) = z E\in\mathfrak C_0'$.

\section{Conclusions and outlook}
\label{conclusions}
\label{outlook}
\label{parameterization}

A standard way to describe meromorphic connections and their deformations is to consider the short exact sequence of sheaves 
\begin{eqnarray}
0 \, \rightarrow \, \underline{\ad\mathcal P} \, \hookrightarrow \, \ad\mathcal P \,  \overset\nabla\rightarrow \, \ad\mathcal P\otimes K(D)\rightarrow \, 0,
\end{eqnarray}
where $\underline{\ad\mathcal P} \hookrightarrow \ad\mathcal P$ denotes the injection of the sheaf of $\nabla$-flat sections of $\ad\mathcal P$ into all of them. It yields the following exact sequence of \v{C}ech cohomology groups over $\mathcal C_{\bold p}$
\begin{small}\begin{eqnarray}
\label{CechDeformations}
0  \rightarrow  \Ho^0(\underline{\ad\mathcal P}) \rightarrow  \Ho^0(\ad\mathcal P) \rightarrow \qquad \qquad\qquad  \qquad  \qquad\qquad \quad\qquad \qquad\qquad\qquad\qquad \nonumber \\
 \qquad\qquad \qquad  \Ho^0(\ad\mathcal P\otimes K(D)) \rightarrow  \Ho^1(\underline{\ad\mathcal P}) \rightarrow  \Ho^1(\ad\mathcal P)   \qquad\qquad \qquad\qquad\qquad\qquad \ \, \\
 \qquad \qquad\qquad  \qquad  \qquad\qquad \qquad \qquad    \rightarrow \Ho^1(\ad\mathcal P\otimes K(D))
\rightarrow  \Ho^2(\underline{\ad\mathcal P}) \rightarrow 0,  \nonumber
\end{eqnarray}\end{small}
symmetric with respect to its centre $\Ho^1(\underline{\ad\mathcal P})$ by Serre duality \cite{Har1977}. This central term encodes simultaneous deformations of the meromorphic connection $\nabla$ and complex structure on the underlying bundle. In general, it does not split into the direct sum of the bundle complex structure deformations $\Ho^0(\ad\mathcal P\otimes K(D))$ (ie. so called \textit{Higgs fields}) and the deformations $\Ho^1(\ad\mathcal P)$ of the connection. The two cohomology groups in the first (and dually last) line of (\ref{CechDeformations}) are easily interpreted as the intersection of the commutants of the monodromy and Stokes matrices of $\nabla$, and as the centre of the Lie algebra $\Lieg$ respectively.

This classical \textit{cohomological} picture naturally describes the deformations of pairs $(\mathcal P,\nabla)$ of principal bundle equipped with meromorphic connections. The deformations of the latter being encoded by classes of cocycles in $\Ho^1(\ad\mathcal P)$. In this paper, we have supplemented it with a dual \textit{homological} picture in terms of quantum geometry of connections, where these deformations are described as in \cite{BER2018, BeEy2019} by classes of cycles in $\widehat\Ho_1=\Ho_1(\underline{\ad\mathcal P})$, with \textit{flat} local coefficients, the connections themselves being described as the collections of their multi-valued flat sections of $\ad\mathcal P$, or equivalently as maps $\mathfrak C_0'\longrightarrow \ad\mathcal P$ associating them to localised quantum divisors, namely divisors on $\mathcal C_{\bold p}$ with local coefficients in $\ad\mathcal P$ that are supported at a single base-point. This map was constructed using the canonical holomorphic connections of \cite{BiHu2021}, as well as insights from the envelope point of view on RG \cite{Kun1995}.

As stated in Section \ref{solutions} however, those structural results on families of meromorphic equivariant connections over Riemann surfaces relied on the assumption that we already had a fundamental adjoint solution at our disposal. This makes the integral representations (\ref{IntegralForm}) the perfect gateway for a quantum geometric description of wild character varieties from a semi-classical point of view. The exact WKB method indeed being expected to provide such solutions canonically \cite{Dun1932, BNR1982, Vor1983, GMN20XX, Tak2016, ABS2019} , and which the author of the present paper is currently developing.

\section*{Acknowledgements}

The author would like to thank Sanika Diwanji for support, as well as Bertrand Eynard, Jacques Hurtubise, and Reinier Kramer for discussions. This work was also supported by a PIMS-CNRS Postdoctoral Fellowship as well as the Natural Sciences and Engineering Research Council. The research and findings may not reflect the views of these institutes. The University of Alberta respectfully acknowledges that we are situated on Treaty 6 territory, traditional lands of First Nations and M\'etis people.

%\pagebreak

\bibliographystyle{morder5}

\begin{footnotesize}

\end{footnotesize}

\end{document}